\documentclass[aps,prl,twocolumn,groupedaddress]{revtex4-1}
\usepackage{graphicx,psfrag}
\usepackage{graphicx}
\usepackage{natbib}
\usepackage{amsmath}                         
\usepackage{amssymb}                         
\usepackage{amsfonts} 
\usepackage{times}                        
\usepackage{labelfig}

\usepackage[usenames,dvipsnames]{color}

\newcommand\bx{\mathbf{x}}

\begin{document}

\title{Genesis of Streamwise-Localized solutions from Globally Periodic Travelling Waves in Pipe Flow }

\author{M. Chantry$^1$}
\email{Matthew.Chantry@bris.ac.uk}
\author{A. P. Willis$^{2}$}
\email{A.P.Willis@shef.ac.uk}
\author{R. R. Kerswell$^1$}
\email{R.R.Kerswell@bris.ac.uk }

\affiliation{$^1$ School of Mathematics, University of Bristol, Bristol BS8 1TW, UK.\\ 
$^2$ School of Mathematics and Statistics, University of Sheffield, Sheffield S3 7RH, UK.}

\date{\today}

\begin{abstract}
The aim in the dynamical systems approach to transitional turbulence is 
to construct a scaffold in phase space for the dynamics using simple 
invariant sets (exact solutions) and their stable and unstable 
manifolds. In large (realistic) domains where turbulence can co-exist 
with laminar flow, this requires identifying exact localized solutions. 
In wall-bounded shear flows the first of these has recently been found 
in pipe flow, but questions remain as to how they are connected to the 
many known streamwise-periodic solutions. Here we demonstrate 
that the origin of the first localized solution 
is in a modulational symmetry-breaking Hopf 
bifurcation from a known global travelling wave that has 2-fold 
rotational symmetry about the pipe axis. Similar behaviour is found for 
a global wave of 3-fold rotational symmetry, this time leading to two 
localized relative periodic orbits. The clear implication is that many 
global solutions should be expected to lead to more realistic localized 
counterparts through such bifurcations, which provides a constructive 
route for their generation. 
\end{abstract}


\maketitle

The transition to turbulence of wall-bounded shear flows remains a 
fundamental problem in fluid mechanics because of the complexity of the 
flows observed, the depth of the mathematical issues involved and the 
numerous practical implications. The central issue is that shear flows 
are usually linearly stable (e.g. pipe flow and plane Couette flow) yet 
nonlinearly unstable to small but finite-amplitude disturbances that 
typically trigger transition in practical situations. This transition is 
abrupt, immediately leading to complicated spatiotemporal flows 
\cite{reynolds1883,emmons51,bottin98}. There has, however, been 
considerable recent success in viewing such transitional fluid flows as 
large dynamical systems where transition amounts to the flow being 
disturbed out of the basin of attraction of the laminar shear flow 
\cite{kerswell05,eckhardt07,kawahara12}. Key to making progress with 
this picture has been the discovery of simple exact unstable solutions 
(equilibria, travelling waves, periodic orbits) to the Navier-Stokes 
equations \cite{nagata90,waleffe98,faisst03,wedin04,duguet08}, which are born in 
saddle node bifurcations. These solutions are either embedded in the basin 
boundary (between the laminar and turbulent states) or 
sit in the basin of attraction of the turbulent state, possibly even being buried in the turbulent attractor itself.
If the turbulent state is not an attractor,
the more general concept of an `edge' (a codimension one hypersurface dividing initial conditions that enjoy a turbulent episode from those that immediately relaminarise) is needed.
Mapping out the stable and unstable 
manifolds of some of these solutions embedded in the 
basin boundary (or edge) has revealed 
much about what particular disturbances are most efficient in triggering transition 
\cite{pringle10, monokrousos11, pringle12} and the transition process 
itself \cite{skufca06, schneider07}. Doing the same for 
solutions embedded in the turbulent attractor generates a skeleton in 
phase space over which the turbulent dynamics is draped \cite{gibson08,willis12}.

Most of this progress has been in the context of shear flows studied 
over artificially small periodic domains (for theoretical convenience) 
where laminar and turbulent states do not co-exist and all the known 
exact solutions are spatially global. In real shear flows, however, 
transition is a spatiotemporal phenomenon where patches of turbulence 
are triggered that coexist with the laminar state 
\citep[e.g.][]{reynolds1883, emmons51, bottin98}. The outstanding challenge is 
therefore to extend this dynamical systems approach to larger domains, 
which naturally focuses attention on identifying localized exact 
solutions.

The first step in this direction was taken in plane Couette flow where 
\textit{spanwise}-localized steady states and travelling waves were found 
\cite{schneider10localized} using an edge-tracking technique 
\cite{itano01,skufca06,schneider07}.  This technique identifies a 
relative attractor on the edge and requires it, atypically,
 to be a simple rather than chaotic state. Perhaps more important 
than discovering the states themselves was the realization (after 
continuing them in parameter space) that they originated in bifurcations 
from known global solutions, periodic in the spanwise-direction 
\cite{schneider10snakes}. Between the local and global solutions 
in parameter space is a regime where the solution branch folds back and 
forward adding repetitions of 
the localized solution, termed homoclinic snaking \cite{knobloch08}.
This suggested that further (at least) spanwise-localized versions of the 
catalogue of known solutions in small domains would quickly be isolated, 
but so far these have failed to materialize, although some 
progress is being made using informed guesswork \cite{gibson13}. 
To tackle the harder problem of \textit{streamwise}-localization 
(the large streamwise flow strongly couples repetitions of
the wave),
the same edge-tracking technique
has been employed in pipe flow to discover streamwise localisation \cite{duguet10}, where 
quasi-periodic behavior was observed, and recently the first streamwise-localised simple 
periodic solution in shear flows \cite{avila13}.  More precisely, this is a relative periodic orbit 
(RPO) where the velocity field at a period later is related by a spatial 
shift to the initial flow. In this case no snaking behaviour was found 
leaving the connection back to a global state unclear.

In this Letter we discover this connection by demonstrating how the 
\textit{streamwise}-localized solution \cite{avila13} arises through a 
symmetry-breaking bifurcation from a globally periodic state. This is 
important because it shows how to constructively generate further 
localized solutions from the numerous global states known, rather than 
relying on edge tracking. 
(This technique has been invaluable, particularly in small domains, 
but becomes less successful at finding exact solutions as dimension size 
increases e.g.\ with pipe length \cite{mellibovsky2009}.) To establish 
this connection we have traced the RPO \cite{avila13} 
towards decreasing pipe length and found a bifurcation from the N2 
family of highly-symmetric travelling wave solutions \cite{pringle09}. 
To investigate the prevalence of the mechanism, we show how exactly the same 
type of bifurcation gives rise to another (new) streamwise-localized RPO 
from N3, the cousin of the global travelling wave N2 with 3-fold rather 
than 2-fold rotational symmetry.

%
%
\begin{figure}
\centerline{\includegraphics[width=0.97\columnwidth]{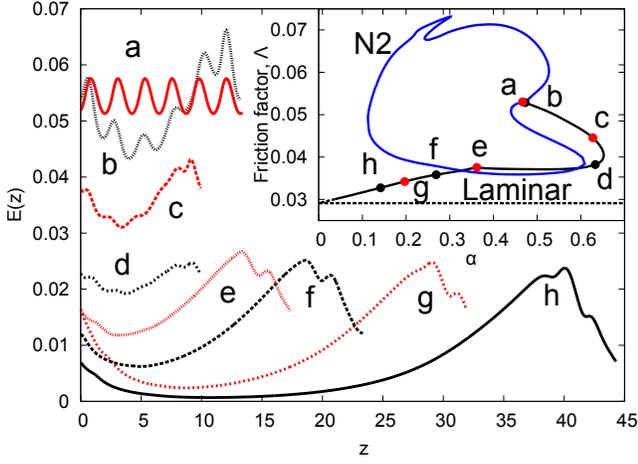}}
\caption{Localization of the bifurcation from the N2 travelling wave (three copies in $z$). Main plot: $E(z):=\int_0^{2\pi} d\theta \int_0^{1} sds \tfrac{1}{2}\mathbf{u}^2$ demonstrating localization along the continuation curve. Inset, continuation in $\alpha=\tfrac{2\pi}{L}$ against friction factor, $\Lambda$ (the axial pressure gradient averaged over the whole pipe and multiplied by $R/\rho U^2$, where $R$ is the pipe radius and $U$ the mean axial flow speed). The branch moves towards smaller domains before turning in a saddle node bifurcation and localizing. The friction factor's linear dependence upon $\alpha$ signals localization.
}
\label{fig:contR2}
\end{figure}

Computations were carried out using a hybrid spectral finite-difference 
code \cite{willis09} with a resolution of 16 Fourier modes in the 
azimuthal direction (so capturing up to the 32nd and 48th wavenumbers in 
the 2-fold and 3-fold rotational calculations respectively), 64 finite 
difference points in the radial direction, and 4 Fourier modes per 
radial length in the streamwise direction (results were checked for 
robustness by both increasing spatial and temporal resolution by a 
factor of approximately $\sqrt{2}$). The localized solution of 
\citet{avila13} was rediscovered using the edge-tracking technique 
\cite{itano01,skufca06} at $Re=2200$ in a domain of length 50 pipe radii 
(hereon 50R). The following rotational and reflectional symmetries were
imposed \begin{align} \mathbf{R}_m:& 
\left(u,v,w\right)\left(s,\theta,z\right) \rightarrow \left(u,v,w\right) 
\left(s,\theta + \tfrac{2\pi}{m},z\right), \\ \mathbf{Z}:& 
\left(u,v,w\right) \left(s,\theta,z\right) \rightarrow 
\left(u,-v,w\right)\left(s,-\theta,z\right), \end{align} where 
$\left(s,\theta,z\right)$ are the usual cylindrical coordinates, 
$\mathbf{u}=\left(u,v,w\right)$ are the corresponding velocity 
components and $m$ is the order of the rotational symmetry subspace 
($m=2$ here). Convergence and continuation of solutions was achieved 
using a Newton-Krylov-hookstep algorithm \cite{viswanath07}.  The 
highly-symmetric travelling wave N2 \cite{pringle09} has the extra 
shift-\&-rotate symmetry: \begin{equation} \mathbf{\Omega}_m: 
\left(u,v,w\right)\left(s,\theta,z\right) \rightarrow 
\left(u,v,w\right)\left(s,\theta+ \tfrac{\pi}{m},z- 
\tfrac{\pi}{\alpha}\right) \end{equation} where $\tfrac{2\pi}{\alpha}$ 
is the wavelength of the solution.

%
%
%

For clarity, we shall describe the bifurcation moving from the 
travelling wave N2 towards the localized RPO of \cite{avila13}: see 
figure \ref{fig:contR2}. The localized branch bifurcates as a 
modulational symmetry-breaking Hopf bifurcation from the N2 solution 
branch \cite{pringle09} at $\alpha_w\approx 1.41$.  The bifurcation 
breaks the $\mathbf{\Omega}_2$ symmetry group on three copies of the 
underlying solution leaving a bifurcated solution with a smallest 
wavenumber of $\alpha=\alpha_w/3\approx 0.47$.

To understand the effect of breaking a symmetry in a Hopf bifurcation we 
consider a symmetry $C$ of order $N$. An eigenfunction, 
$\mathbf{u}_1$, breaking this symmetry satisfies 
$ C\mathbf{u}_1 = e^{2\pi i n/ N }\mathbf{u}_1$
where $n$ is a divisor of $N$. The new bifurcated branch can be 
written as 
%
$
\mathbf{u}\left(t\right) = \sum_{p=-\infty}^{\infty} \epsilon^p \mathbf{u}_p e^{ip\omega t}, 
$%
where $C \mathbf{u}_0 = \mathbf{u}_0$ is the underlying 
solution, $\epsilon$ measures the distance along the branch from the 
bifurcation point and $\omega$ is the Hopf frequency. From the quadratic 
nonlinearity of the Navier-Stokes equations, the higher temporal 
harmonics obey
$C \mathbf{u}_p = e^{2\pi i n p / N 
}\mathbf{u}_p$
so that in the Galilean frame $z^*=z-ct$, 
where $c$ is the wavespeed of the travelling wave 
$\mathbf{u}_0(s,\theta,z^*)$, 
%

\begin{align} 
\mathbf{u}\left(\bx,t_0+\frac{2\pi n}{\omega N}\right) &= \sum_{p=-\infty}^{\infty} \! \! \mathbf{u}_p e^{2\pi i n p/N 
}e^{i p \omega t_0} \nonumber\\ 
&= \sum_{p=-\infty}^{\infty} \! \!C\mathbf{u}_p e^{ip\omega t_0} = C 
\mathbf{u}\left(\bx,t_0\right). 
\end{align}
This means that the 
pointwise-in-time symmetry $C$ is broken down to a weaker spatiotemporal 
symmetry. In our specific case, $C=\Omega_2$, which has order $N=6$ over 
3 wavelengths of $\mathbf{u}_0$, is broken in a Hopf bifurcation with 
$n=1$ down to the spatiotemporal symmetry 
\begin{multline} 
\mathbf{\Omega}_{m,\tfrac{nT}{N}} \!:\! 
\mathbf{u}\left(\theta,z^*,t\right) \!\rightarrow\!  
\mathbf{u}\left(\theta+\tfrac{\pi}{m},z^*\! \!- \! 
\!\tfrac{\pi}{\alpha_w},t+\tfrac{nT}{N}\right), \label{eqn:OmT} 
\end{multline} where $T=2\pi/\omega$ (and $s$ dependence has been 
suppressed). This realisation is actually crucial in making the 
calculations possible since the RPO needs only to be traced over a time 
$\tau:=T/6$ rather than the full period $T$ making both the 
time-stepping calculations 6 times quicker and the 
Newton-Krylov-hookstep algorithm more likely to converge.

Continuing the RPO branch away from the bifurcation where $\tau=14.3$ (figure \ref{fig:contR2} inset) 
shows that the kinetic energy, $E(z)$, initially increases in one part of the domain 
and decreases in the other (curve b, figure \ref{fig:contR2}) but then generally decreases
 everywhere as the domain shrinks (curve c) and then lengthens (curves d-h) with $\tau$ asymptoting to $29.8$.
Examining slices of the velocity field indicates that the roll-streak structure around the energetic peak 
of the solution changes little from point (e) onwards but $E(z)$ near the energetic peak continues to 
evolve and the energetic  minimum is far from zero indicating that localisation is not yet complete. For  $\alpha \leq 0.2$, however, the friction factor has a linear dependence 
upon $\alpha$ intercepting the laminar value ($\Lambda=\tfrac{64}{Re}$) 
at $\alpha=0$. This linear regime indicates invariance of the solution 
to the domain length: the pressure gradient across the localised
solution remains 
constant and therefore the pressure gradient averaged over the full domain 
scales linearly with $\alpha$.

%
%

%
%
\begin{figure}
\centerline{\includegraphics[width=0.95\columnwidth]{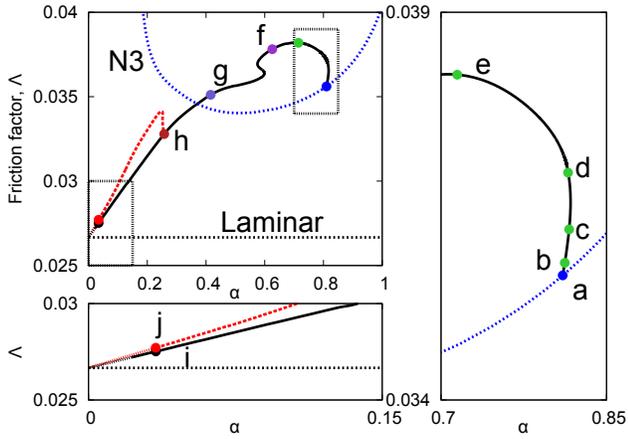}}
\caption{Continuation of the bifurcation from the N3 travelling wave (three copies in $z$) in $\alpha$ plotted against friction factor, $\Lambda$. Curve initially bifurcates towards smaller domains before turning in a saddle node; after this the solution begins to localize. At $\alpha \approx 0.26$ a secondary bifurcation breaks the $\mathbf{\Omega}_{3,T/6}$ symmetry leading to a second localized solution (plotted in red/dashed). Dots (a)-(h) correspond to solutions plotted in figure \ref{fig:shapeR3} and (i), (j) to those plotted in figure \ref{fig:local}. Inset right, zoomed into region near bifurcation. Inset below, zoomed into $\alpha < 0.15$ region demonstrating linear behaviour as domain length tends to infinity ($\alpha \rightarrow 0$).}
\label{fig:contR3}
\end{figure}
%
%
\begin{figure}
\centerline{\includegraphics[width=0.95\columnwidth]{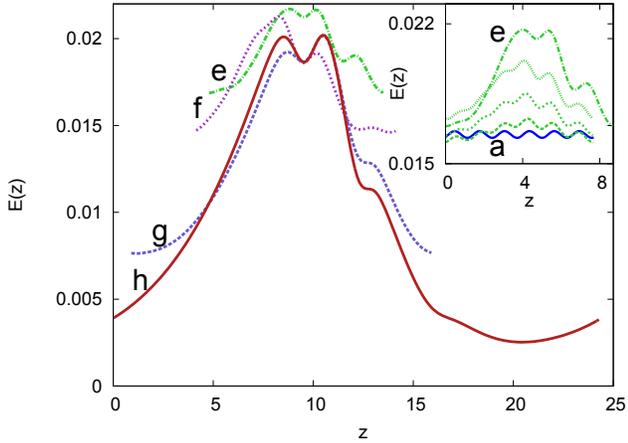}}
\caption{$E(z)$ at dots along the continuation curve (figure \ref{fig:contR3}). Inset, behaviour close to the bifurcation as the solution increases in amplitude. Main, further continuation up to the secondary bifurcation as the solution localizes. Solutions (e)-(h) have been aligned to demonstrate solution evolution along the continuation curve. Solution (e) is plotted in both main \& inset for comparison. Note $E(z)$ has two wavelengths to every wavelength of N3 due to the $\Omega_2$ symmetry (see solution (a)). }
\label{fig:shapeR3}
\end{figure}

To examine whether this type of connection is generic,
edge-tracking was 
also carried out at a similar Reynolds number and domain size 
($Re=2400$, $L=50R$) but in a different rotational symmetry subspace, 
$\mathbf{R}_3$ (so $m=3$ now), while again enforcing $\mathbf{Z}$ 
symmetry. A localized RPO attractor within the edge is found, 
which can also be continued to smaller domains ending at the global periodic 
travelling wave N3: see figure \ref{fig:contR3}. As before, the bifurcation 
off the N3 solution is a modulational symmetry-breaking Hopf 
bifurcation, on three copies of the solution, with symmetry group
$\left<\mathbf{R}_3, \mathbf{Z}, \mathbf{\Omega}_3\right>$ at 
$\alpha_w\approx 2.43$. The $\mathbf{\Omega}_3$ symmetry is broken, 
leading to the time dependent symmetry (\ref{eqn:OmT}) for $m=3$, $n=1$ 
with an initial period of $\tau=\tfrac{T}{6}=37.5$. The curve initially 
bifurcates towards smaller domains before rapidly turning in a 
saddle-node bifurcation. 

Figure \ref{fig:shapeR3} demonstrates the 
rapidly changing shape of the solution during this period as $E(z)$ 
quickly begins to show localization. At point (h) a secondary period doubling bifurcation occurs, 
breaking $\mathbf{\Omega}_{3,T/6}$. Following this new branch (red 
curve) leads to the localized edge state. The original solution branch 
fully localizes but has an addition complex conjugate pair of unstable 
eigenvalues and is therefore not an attractor within the edge. Both 
solution curves enter a regime with linear dependence of friction factor 
on $\alpha$ as $\alpha \rightarrow 0$ indicating localization 
(with $\tau \rightarrow 32.9$ and $T \rightarrow 66.1$ for symmetry-maintained 
and symmetry-broken RPOs respectively).

%
%
\begin{figure}
\SetLabels 
(+0.07*0.9){\large (a)} \\
\endSetLabels 
\leavevmode
\strut\AffixLabels{\centerline{\includegraphics[width=0.95\columnwidth]{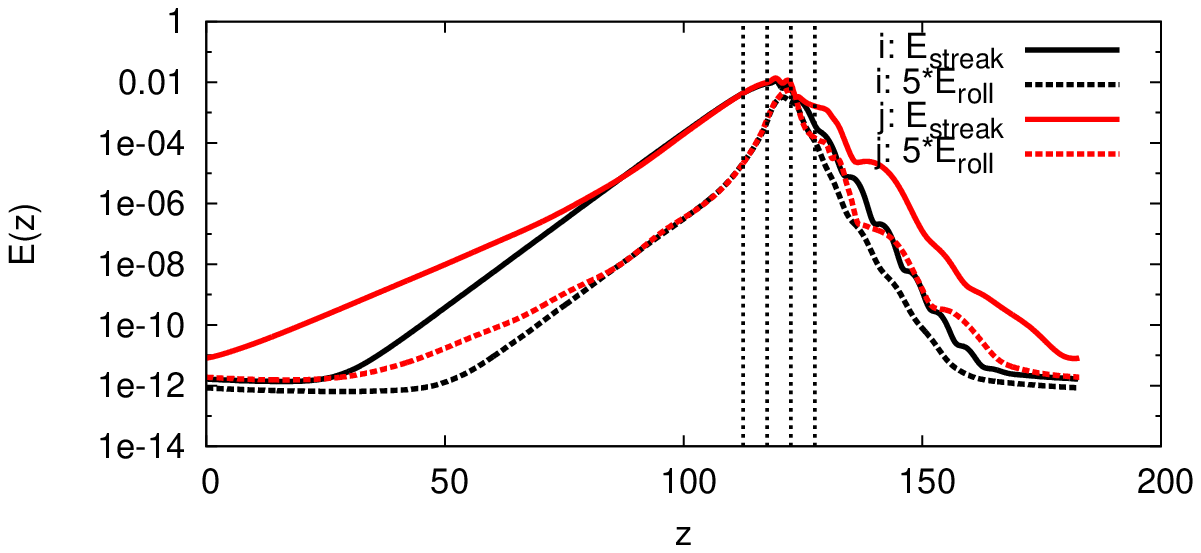}}}
\\
\SetLabels 
(+0.07*0.9){\large (b)} \\
\endSetLabels 
\leavevmode
\strut\AffixLabels{\centerline{\includegraphics[width=0.95\columnwidth]{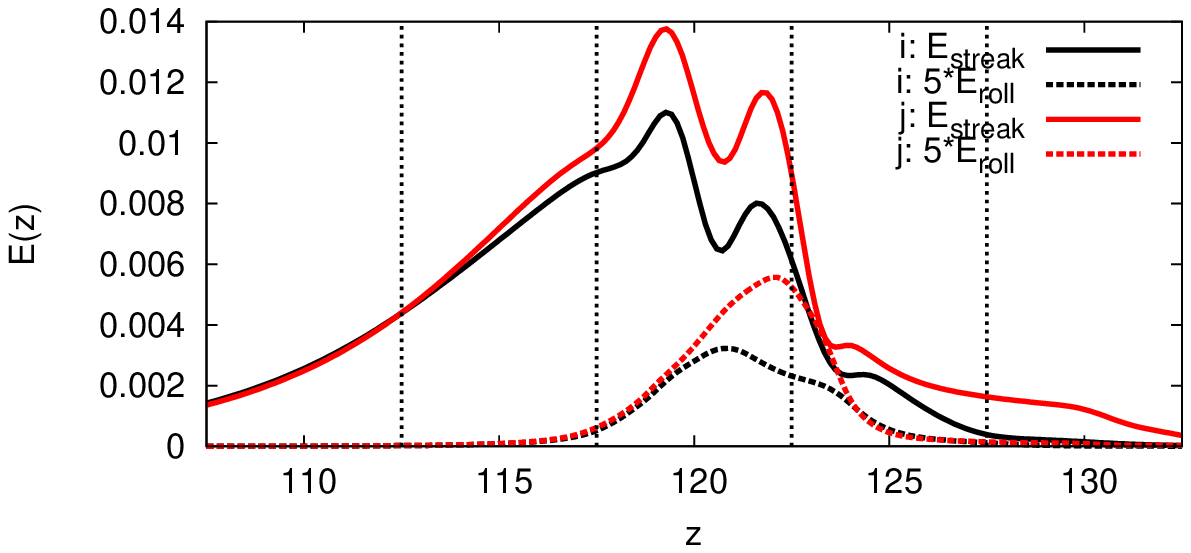}}}
\caption{Energy in streaks ($\theta$-dependent streamwise flow), and rolls ($\theta$-dependent cross-stream flow) as a function of downstream position, $z$, for the two $\mathbf{R}_3$-symmetric localized solutions: the $\mathbf{\Omega}_{3,T/6}$ symmetric solution, labelled (i); and symmetry broken solution, labelled (j). Dotted vertical lines indicate slice locations for velocity visualization plotted in figure \ref{fig:slice}. Figure (a) Logarithm of energy, the $\mathbf{\Omega}_{3,T/6}$ symmetric solution spans approximately $120R$ including exponential tails while the symmetry-broken solution has length of approximately $180R$. Figure (b) Focused onto the centre of the domain, the amplitude of the solutions differ but have very similar roll and streak structure (not plotted).
}
\label{fig:local}
\end{figure}

%
%
\begin{figure}
\begin{center}
\SetLabels 
(+0.07*0.9){\large (a)} \\
\endSetLabels 
\leavevmode
\strut\AffixLabels{\includegraphics[width=0.45\columnwidth]{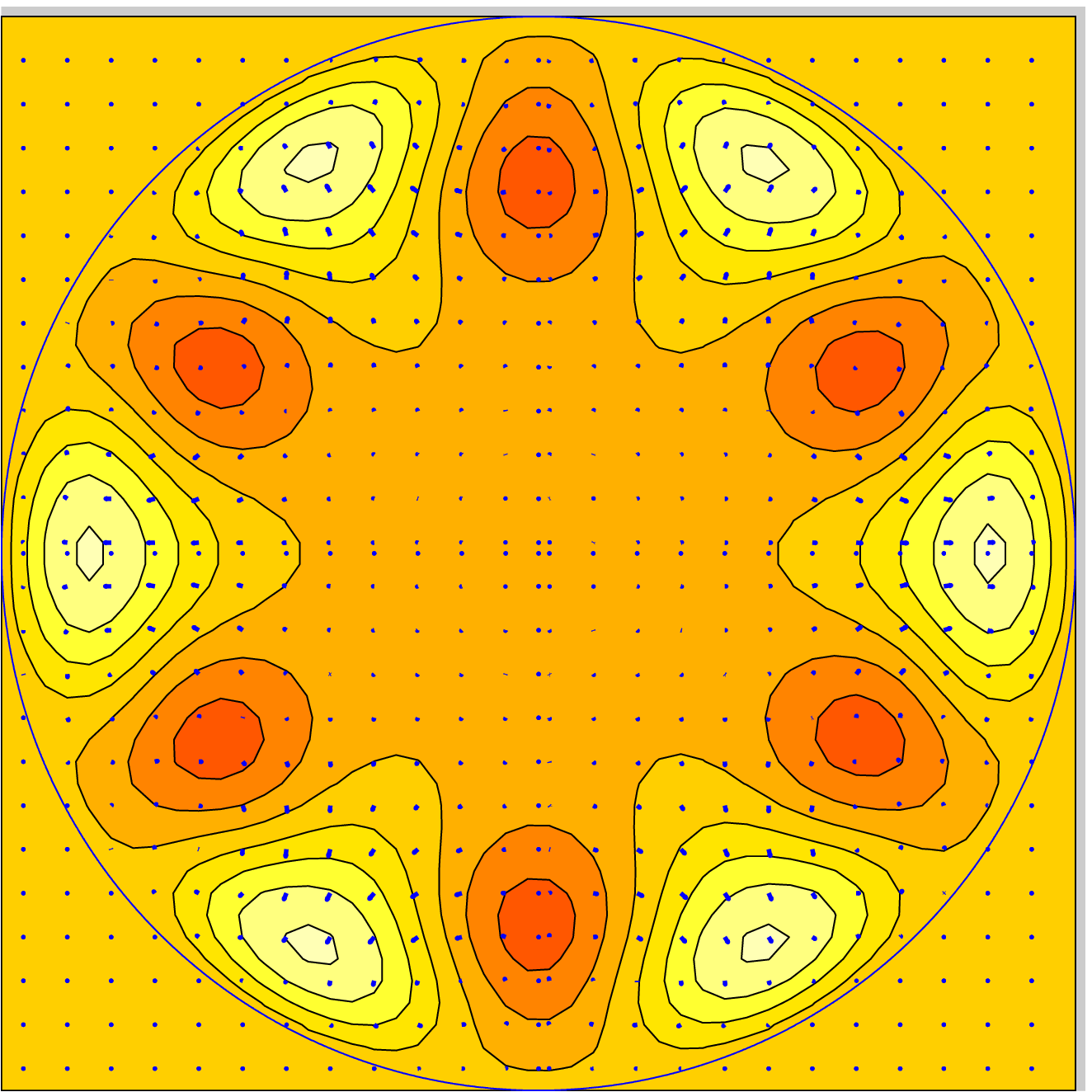}}
\SetLabels 
(+0.07*0.9){\large (b)} \\
\endSetLabels 
\leavevmode
\strut\AffixLabels{\includegraphics[width=0.45\columnwidth]{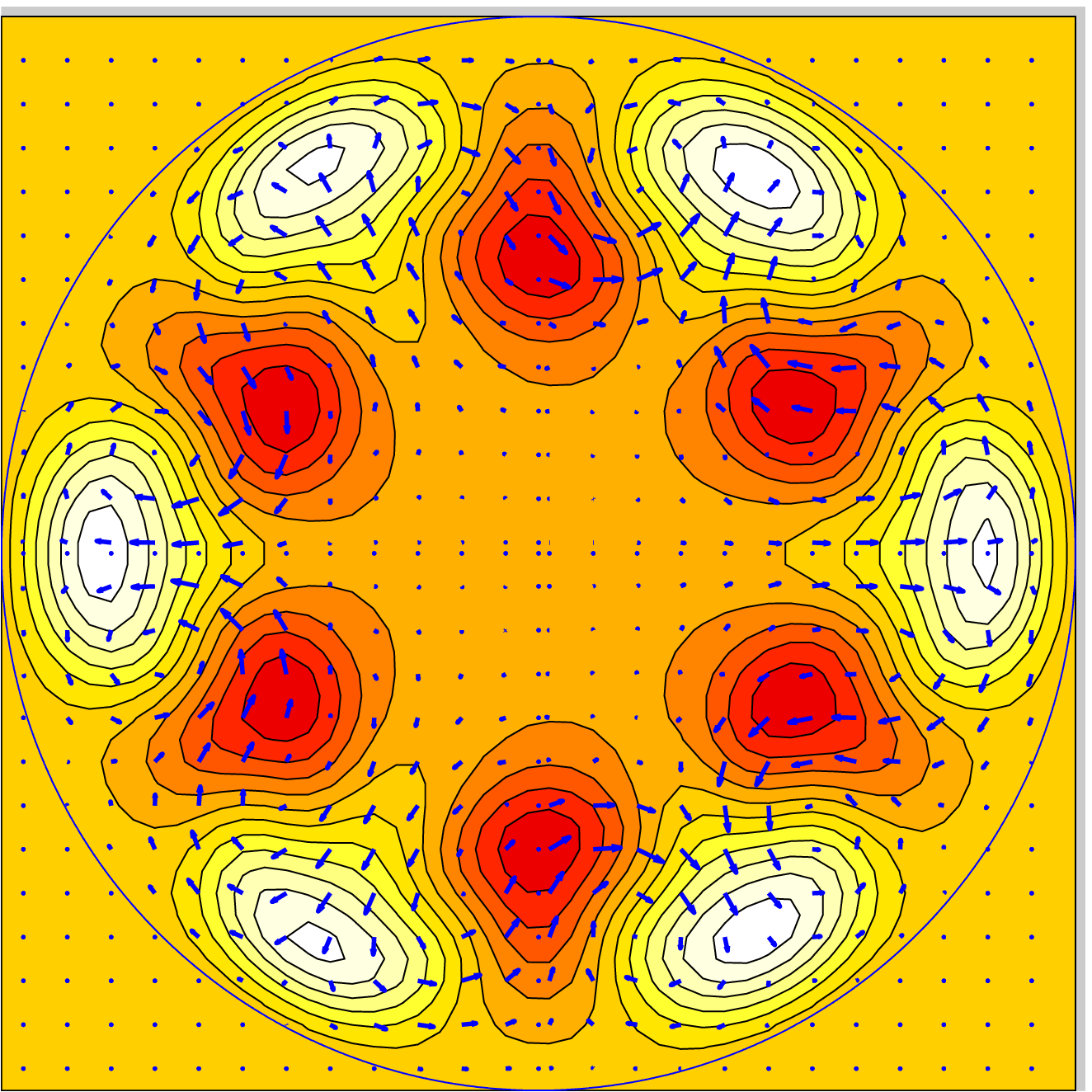}}
\\
\SetLabels 
(+0.07*0.9){\large (c)} \\
\endSetLabels 
\leavevmode
\strut\AffixLabels{\includegraphics[width=0.45\columnwidth]{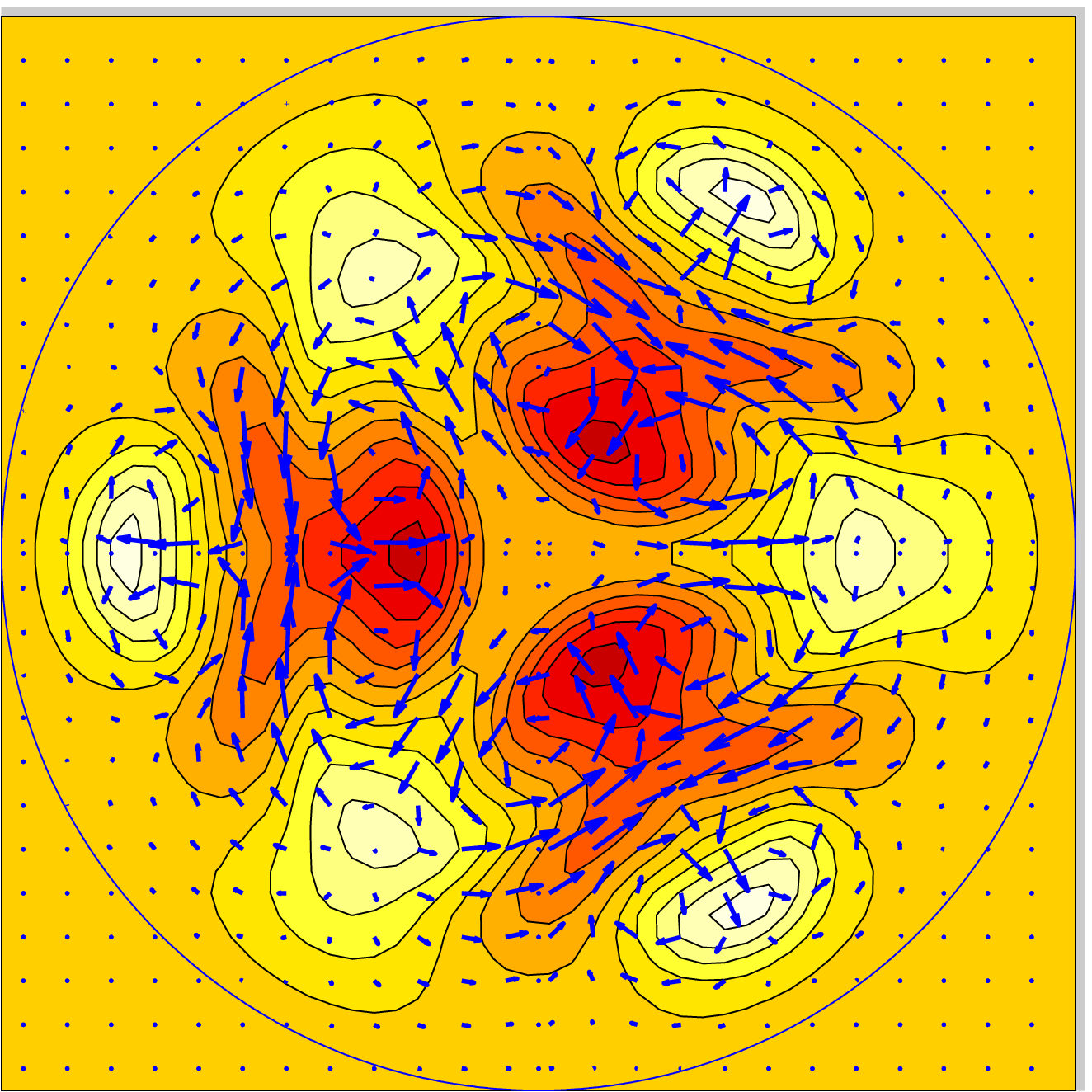}}
\SetLabels 
(+0.07*0.9){\large (d)} \\
\endSetLabels 
\leavevmode
\strut\AffixLabels{\includegraphics[width=0.45\columnwidth]{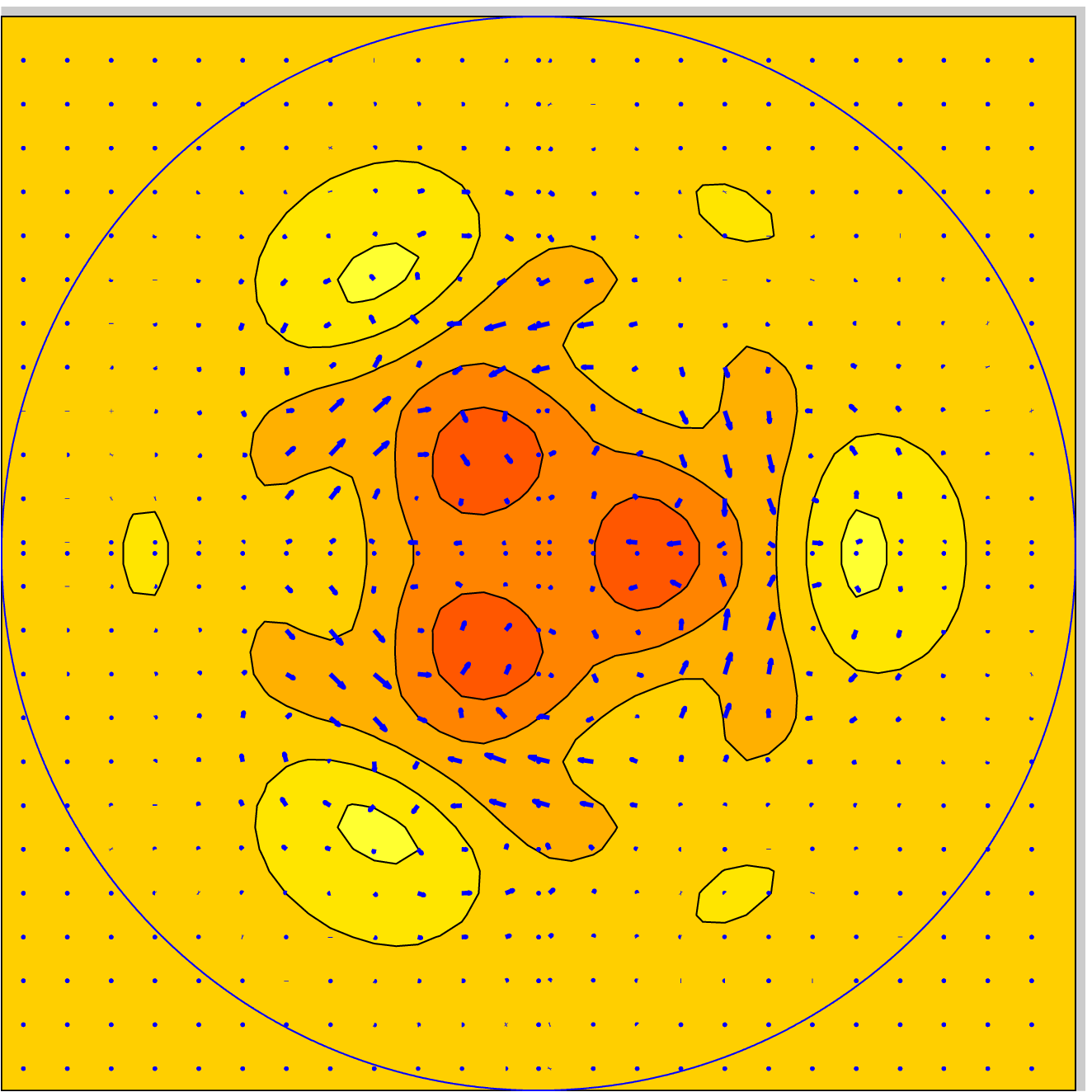}}
\end{center}
\caption{Flow (deviation from laminar state) at slices indicated by dotted lines in figure \ref{fig:local} for solution (i). Downstream flow plotted using contours with white for relatively fast flow, and red for slow flow. 
Arrows indicate cross-stream flow.
}
\label{fig:slice}
\end{figure}

Formally, the solutions can only really be considered fully localized if 
their extremities reach the error tolerance of the code, 
$O\left(10^{-13}\right)$. To examine this, we took the 
$\mathbf{\Omega}_{3,T/6}$ symmetric localized branch and continued to 
$\alpha = 0.035$ ($L=180R$). In figure \ref{fig:local} we plot the 
energy as a function of $z$ for the parts of the flow representing the 
rolls and the streaks at a point along the orbit. The 
$\mathbf{\Omega}_{3,T/6}$ symmetric solution, labelled (i), demonstrates 
the existence of exponential tails both upstream and downstream of the 
localized solution. The actual length of the solution can be viewed as 
approximately $120R$ with similar spatial decay rates for both rolls and 
streaks. The second localized solution within the $\mathbf{R}_3$ 
subspace, which is an attracting edge state, is also plotted (labelled 
(j)). While $E(z)$ differs between these solutions, the streak and roll 
structure of the two solutions is extremely similar. Similarities 
between the streak structure of these solutions and that of the lower 
branch N3 solution (figure 4(b) of \cite{pringle09}) can be observed in 
figure \ref{fig:slice}, with an interior wavelength corresponding to the 
N3 solution at $\alpha=1.25$.

%
%

In this work we have shown how the streamwise-localized RPO of 
\cite{avila13} is connected to a global highly-symmetric travelling wave 
N2 \cite{pringle09} via a simple modulational Hopf bifurcation with no 
snaking (i.e. successive wavelengths of the global state do not 
disappear at saddle node bifurcations). A further streamwise-localized 
RPO has been found to arise by exactly the same mechanism from N3 and 
yet another (more stable) RPO has been identified that bifurcates 
off this RPO. The implication seems clear that many more localized 
states undoubtedly originate via these modulational bifurcations from 
the wealth of spatially global states now known in shear flows. Most 
significantly, our results indicate that the modulational wavelength may 
only be 3 times longer than that of the global state. However, 
identifying and tracking modulational bifurcations is not guaranteed to 
ultimately produce a localized state. By reverse engineering, the two 
examples described here lead to localization but this is an inherently 
nonlinear behaviour that can only be determined by tracking bifurcating 
solution branches. However, this bifurcation approach is at least a {\em constructive} way to 
generate further localized solutions and so represents a systematic technique
to extend the dynamical systems approach to localized turbulence and nicely 
complements the 
edge-tracking approach, which relies on being able to select a $Re$ or 
symmetry subspace to manufacture a simple localized edge state. 
Finally, it's worth remarking that there now seems no reason why a steady 
modulational instability might not give rise to a {\em steady} 
streamwise-localised travelling wave.

\bibliography{ref}

\end{document}